\documentclass[11pt]{article}
\usepackage{main_clean}

\usepackage{background}

\DefineAffil{aff:kapteyn}{Kapteyn Astronomical Institute, University of Groningen, Landleven 12, 9747 AD, Groningen, The Netherlands}
\DefineAffil{aff:icecsic}{Institute of Space Sciences (ICE-CSIC), Carrer de Can Magrans, s/n, 08193 Cerdanyola del Vallès, Barcelona, Spain}
\DefineAffil{aff:leiden}{Leiden Observatory, Leiden University, P.O. Box 9513, 2300 RA, Leiden, The Netherlands}
\DefineAffil{aff:lagrange}{Université Côte d'Azur, Observatoire de la Côte d'Azur, CNRS, Laboratoire Lagrange, France}
\DefineAffil{aff:brera}{INAF -- Osservatorio Astronomico di Brera, via Brera 28, 20121 Milano, Italy}
\DefineAffil{aff:sapienza}{Department of Physics, Sapienza University of Rome, Piazzale Aldo Moro 5, I-00185 Rome, Italy}
\DefineAffil{aff:mpe}{Max Planck Institute for Extraterrestrial Physics, Giessenbachstrasse 1, 85748 Garching, Germany}
\DefineAffil{aff:jodrell}{Jodrell Bank Centre for Astrophysics, Alan Turing Building, University of Manchester, Manchester M13 9PL, UK}
\DefineAffil{aff:mpa}{Max Planck Institute for Astrophysics, Karl-Schwarzschild-Str. 1, 85741 Garching, Germany}
\DefineAffil{aff:oslo}{Institute of Theoretical Astrophysics, University of Oslo, Postboks 1029 Blindern, 0315 Oslo, Norway}
\DefineAffil{aff:oass}{INAF -- Osservatorio di Astrofisica e Scienza dello Spazio, via Piero Gobetti 93/3, 40129 Bologna, Italy}
\DefineAffil{aff:unimore}{Department of Physics, Informatics and Mathematics, University of Modena and Reggio Emilia, 41125 Modena, Italy}
\DefineAffil{aff:iac}{Instituto de Astrofísica de Canarias, 38200 La Laguna, Tenerife, Spain}
\DefineAffil{aff:ull}{Departamento de Astrofísica, Universidad de La Laguna, 38206 La Laguna, Tenerife, Spain}
\DefineAffil{aff:wits}{Wits Centre for Astrophysics, School of Physics, University of the Witwatersrand, PMB 3, 2050 Johannesburg, South Africa}
\DefineAffil{aff:toho}{Department of Physics, Toho University, Funabashi, Chiba 274-8510 Japan}
\DefineAffil{aff:dawn}{DAWN, DTU-Space, Electrovej 327, 2800 Lyngby, Denmark}
\DefineAffil{aff:upenn}{Department of Physics and Astronomy, University of Pennsylvania, Philadelphia, PA 19104, USA}
\DefineAffil{aff:ias}{Université Paris-Saclay, CNRS, Institut d’Astrophysique Spatiale, 91405 Orsay, France}
\DefineAffil{aff:trieste}{Astronomy Unit, Department of Physics, University of Trieste, via Tiepolo 11, 34131 Trieste, Italy}
\DefineAffil{aff:eso}{European Southern Observatory, Karl-Schwarzschild-Str. 2, 85748 Garching bei München, Germany}
\DefineAffil{aff:damtp}{DAMTP, Centre for Mathematical Sciences, Wilberforce Road, Cambridge CB3 0WA, UK}
\DefineAffil{aff:kicc}{Kavli Institute for Cosmology, University of Cambridge, Madingley Road, Cambridge CB3 0HA, UK}
\DefineAffil{aff:dunlap}{Dunlap Dept.\ of Astronomy and Astrophysics, University of Toronto, 50 St George Street, Toronto ON, M5S 3H4, Canada}
\DefineAffil{aff:specola}{Specola Vaticana (Vatican Observatory), V-00120 Vatican City State}

\usepackage{background}

\begin{document}

\begin{titlepage}
\color{white}
\backgroundsetup{ 
    vshift=-40pt,
    scale=1.13,
    angle=0,
    opacity=1,
    contents={\includegraphics[width=\paperwidth,height=\paperheight]{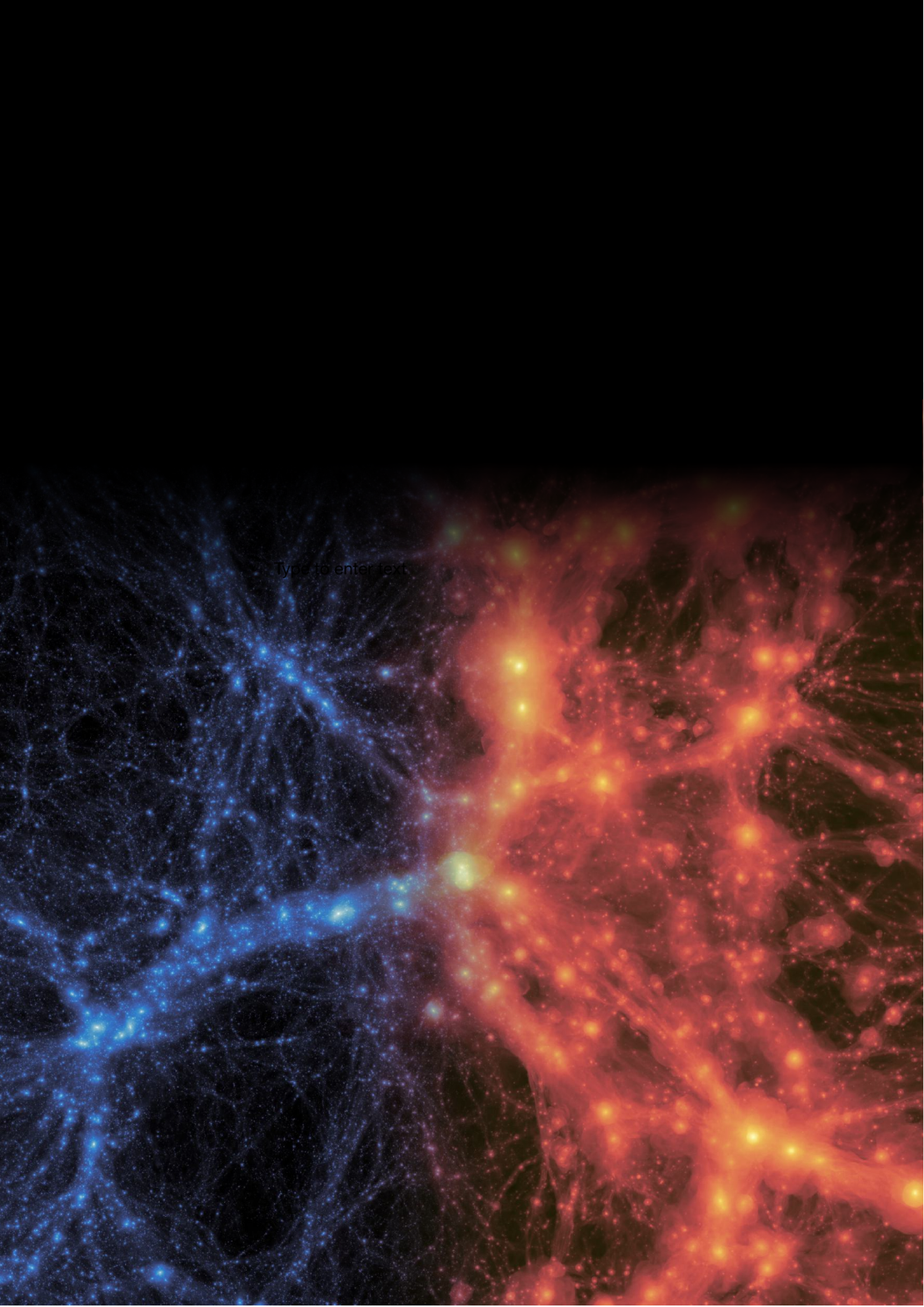}}
}

\textbf{\huge\sffamily
A high-dynamic-range view of the growth\newline
of structure and the warm/hot Universe}
\bigskip

\textbf{\sffamily Authors}

\textbf{Luca Di Mascolo}\affref{aff:kapteyn}\footnote[2]{\color{white}Email: {\hypersetup{urlcolor=paleblue!60!white}\href{mailto:l.di.mascolo@astro.rug.nl}{l.di.mascolo@astro.rug.nl}}}, 
\textbf{Tony Mroczkowski}\affref{aff:icecsic},
\textbf{Joshiwa van Marrewijk}\affref{aff:leiden},
Rémi Adam\affref{aff:lagrange},
Nabila Aghanim\affref{aff:ias},
Stefano Andreon\affref{aff:brera},
Eleonora Barbavara\affref{aff:sapienza},
Elia Stefano Battistelli\affref{aff:sapienza},
Esra Bulbul\affref{aff:mpe},
Jens Chluba\affref{aff:jodrell}
Eugene Churazov\affref{aff:mpa},
Claudia Cicone\affref{aff:oslo},
William Coulton\affref{aff:kicc}\textsuperscript{,}\affref{aff:damtp}
Stefano Ettori\affref{aff:oass},
Massimo Gaspari\affref{aff:unimore},
Ricardo Génova Santos\affref{aff:iac}\textsuperscript{,}\affref{aff:ull},
Matt Hilton\affref{aff:wits},
Adam D.\ Hincks\affref{aff:dunlap}\textsuperscript{,}\affref{aff:specola},
Eelco van Kampen\affref{aff:eso},
Tetsu Kitayama\affref{aff:toho},
Minju Lee\affref{aff:dawn},
John Orlowski-Scherer\affref{aff:upenn},
Charles Romero\affref{aff:upenn},
Laura Salvati\affref{aff:ias},
Alexandro Saro\affref{aff:trieste},
\'{I}\~{n}igo Zubeldia\affref{aff:kicc}\textsuperscript{,}\affref{aff:damtp}

\medskip
\PrintAffiliations
\medskip

\textbf{\sffamily Science keywords}\newline
\textbf{cosmology:} circumgalactic medium, cosmic background radiation, cosmological parameters, dark energy, dark matter, distance scale, large-scale structure of universe, observations; 
{\bf galaxies:} circumgalactic medium, clusters, formation, halos, high-redshift, protoclusters\\

\vfill
{\raggedright \small \it 
\begin{minipage}{.8\textwidth}
        \raggedright
        Simulation of large-scale structure distribution of dark matter (blue) and \newline 
        warm/hot gas as traced by the thermal Sunyaev-Zeldovich effect (orange).\newline
        Data credit: {\hypersetup{urlcolor=paleblue!60!white}\href{https://www.tng-project.org/}{TNG Collaboration}}
    \end{minipage}%
}
\end{titlepage}

\newpage
\backgroundsetup{contents={}}

\begin{tcolorbox}[left=5pt,right=5pt,bottom=2pt,top=2pt,sharp corners,boxrule=0pt,colframe=white,colback=paleblue!20!white]
\textbf{\sffamily Abstract.~} Baryons heat to temperatures above $>\!\!10^5\,\mathrm{K}$ as they accrete onto massive overdensities -- galaxies, groups, clusters, and filaments -- where they ionize and become optically transparent. Deep mm-wave observations such as those with ALMA have begun to probe a handful ($\sim\,$4) of massive systems at $z\!\sim\!2-4$, while low-resolution mm-wave surveys have detected thousands of objects at arcminute resolution out to $z\!\approx\!2$. 
To truly advance the field of the evolution of large-scale structures, mapping the warm/hot distribution of ionized gas out to the redshift of their formation, the ESO community requires a large-aperture single-dish (sub-)mm telescope. This will need to provide several orders of magnitude higher mapping speeds than currently available while preserving the few arcsecond resolution required for imaging the gas and removing contaminating radio and dusty thermal signals across the full (sub-)mm wavelength range.
\end{tcolorbox}

\section{Scientific context and motivation}
Two main classes of observation can be used to infer the history and map the expansion of the Universe after the surface of last scattering ($z\!\!\lesssim\!\!1100$): those relying on distance determinations (e.g., distance ladder, baryon acoustic oscillations, supernovae), and those probing the growth of structure (e.g., galaxy and galaxy cluster counts, lensing and shear maps). As both approaches get to the crux of understanding our cosmic origins, they both must be fully developed. Extensive efforts based on the former strategy are now underway across the visible/near infrared. Here, we focus on the transformational next-generation observations needed for the latter: understanding the growth and evolution of massive structures.

In the standard cosmological framework, the dominant fraction of baryons is in the warm/hot ionized phase \citep{Nicastro2018}, residing within Mpc-scale filaments that connect to the dense nodes of the cosmic web -- galaxy groups and clusters -- and down to the circumgalactic medium surrounding individual galaxies. The physical, thermodynamic, and kinematic properties of the warm/hot baryons provide a direct record of the gravitational and feedback processes that have shaped the growth of structure across cosmic time. Understanding the thermal and dynamical state of the hot, ionized baryons permeating the cosmic web is thus essential for reconstructing the history of structure formation in the Universe.

Even though warm/hot cosmic baryons play a leading role in structure formation, our current observational view remains limited mostly to the local Universe and high-mass end of the halo population. A large fraction of the ionized baryons resides in diffuse, low-surface-brightness environments, where traditional emission-based tracers like X-ray measurements become less effective. The high temperatures further make this phase nearly completely undetectable at optical and near-infrared wavelengths, where the majority of telescopes operate. As a result, vast reservoirs of cosmic baryons have remained beyond the reach of existing facilities. In turn, this has severely limited our ability to fully test our models of structure formation, of the detailed role of feedback processes, and to complete the
\begin{wrapfigure}{r}{0.581\linewidth}
    \centering
    \vspace{-10pt}
    \includegraphics[clip,trim=0 0 0 3pt,width=0.95\linewidth]{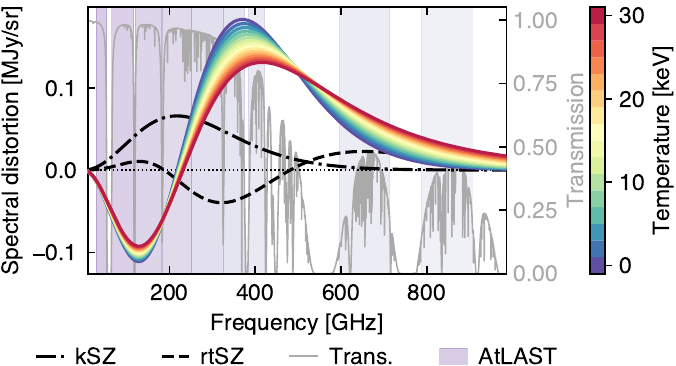}
    \vspace{-6pt}
    \caption{Spectral distortion from to the relativistically corrected thermal SZ effect (coloured solid lines), relativistic SZ correction for $25~\mathrm{keV}$ gas (black dashed line), and kinetic SZ effect (black dot-dashed line). In the background is a comparison of the atmospheric transmission conditions at the Chajnantor Plateau and the proposed spectral bands.\vspace{-20pt}}
    \label{fig:spectra}
\end{wrapfigure} 
census of baryons across cosmic time.\medskip

\textbf{\sffamily SZ view of the warm/hot Universe.~}\linebreak (Sub-)mm wavelengths offer an observational window on this dominant, yet elusive, component of cosmic large-scale structures. This is the spectral range at which the cosmic microwave background (CMB) emission peaks, and acts as a homogeneous backlight to all structures throughout the observable Universe. As the CMB radiation field propagates, its photons are scattered by free energetic electrons in ionized gas reservoirs. The net result is a distinctive distortion of the CMB signal -- the Sunyaev-Zeldovich (SZ) effect.
A defining property of the SZ effect, in contrast to X-ray measurements, is the inherent redshift independence of its surface brightness. Furthermore, the spatial and spectral properties of the SZ effect encode information on the specific velocity distribution of the scattering electrons \cite{Mroczkowski2019}. The dominant SZ term -- the ``thermal'' SZ effect \cite{Sunyaev1972} -- is associated with the electron thermal motion. Its amplitude is a direct proxy of the line-of-sight integral of the electron pressure and, thus, of the total thermal energy of the gas enclosed within cosmic haloes. The bulk peculiar motion of electrons introduces a Doppler shift in the CMB as a second SZ component -- the ``kinetic'' SZ effect \cite{Sunyaev1980}. Both the thermal and kinetic SZ signals further exhibit a direct dependence on the electron temperature, a consequence of the relativistic velocities of the scattering particles. The resulting ``relativistic corrections'' to the thermal and kinetic SZ spectra \citep{Chluba2012} in turn offer a direct observational proxy of the temperature of the ionized gas, without the need for X-ray spectroscopy. The combination of thermal, kinetic, and relativistic SZ effects -- as well as non-thermal SZ terms \citealt{Khabibullin2018,Lee2024} -- can thus be used to gain a comprehensive understanding of the physics and thermodynamics of the ionized gas in large-scale structures, with no redshift limit.\smallskip

\begin{wrapfigure}{r}{0.46\linewidth}
    \vspace{-20pt}
    \includegraphics[width=\linewidth]{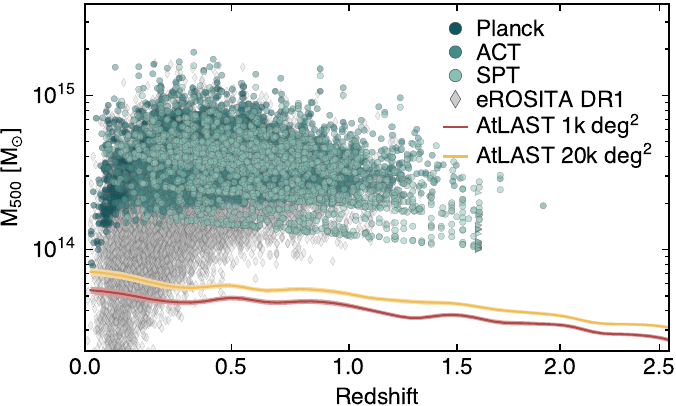}
    \vspace{-23pt}
    \caption{Mass-redshift detection limit for different AtLAST survey strategies \citep{DiMascolo2025} for a fixed survey time of 5 years in comparison with cluster wide-field millimeter (SZ) surveys \cite{Planck2016,Bleem2023,Kornoelje2025,ACT2025} and the eROSITA all-sky (X-ray) survey \cite{Bulbul2024}.}
    \label{fig:survey}
    \vspace{-14pt}
\end{wrapfigure}
\textbf{\sffamily Insights and present-day limitations.} 
Over the past two decades, SZ observations have demonstrated their effectiveness in probing the warm/hot Universe. Wide-area, low-resolution ($\gtrsim$1.5$\arcmin$) SZ surveys have exploited SZ's redshift independence to catalogue $>$10k clusters out to $z\!\!\sim\!\!2$ \cite[Fig.~\ref{fig:survey};][]{Planck2016,Bleem2023,Kornoelje2025,ACT2025}, providing key constraints on the growth of structure and the cosmological parameters that govern it \cite{Miyatake2025}. High-resolution SZ measurements have revealed the detailed pressure structure of clusters, allowing us to begin to decipher the role of, e.g., mergers \citep{Basu2016,DiMascolo2019}, active galactic nuclei \cite{Abdulla2019,OrlowskiScherer2022}, and turbulence \cite{Khatri2016,Romero2023,Adam2025} in injecting energy into and sustaining the intracluster medium. Recently, targeted SZ observations have further pushed into the high-$z$ regime, detecting the hot gas in protoclusters at $z\!\!\gtrsim\!\!2$ \cite{Gobat2019,DiMascolo2023,vanMarrewijk2023,Zhou2025}, as well as revealing the low-density ionized gas in circumgalactic haloes \citep{Liu2025,RiedGuachalla2025} and large-scale filaments \citep{deGraaff2019,Hincks2022}.

\begin{wrapfigure}{r}{0.574\linewidth}
    \vspace{-14pt}
    \includegraphics[clip,trim=0 140pt 0 0,width=\linewidth]{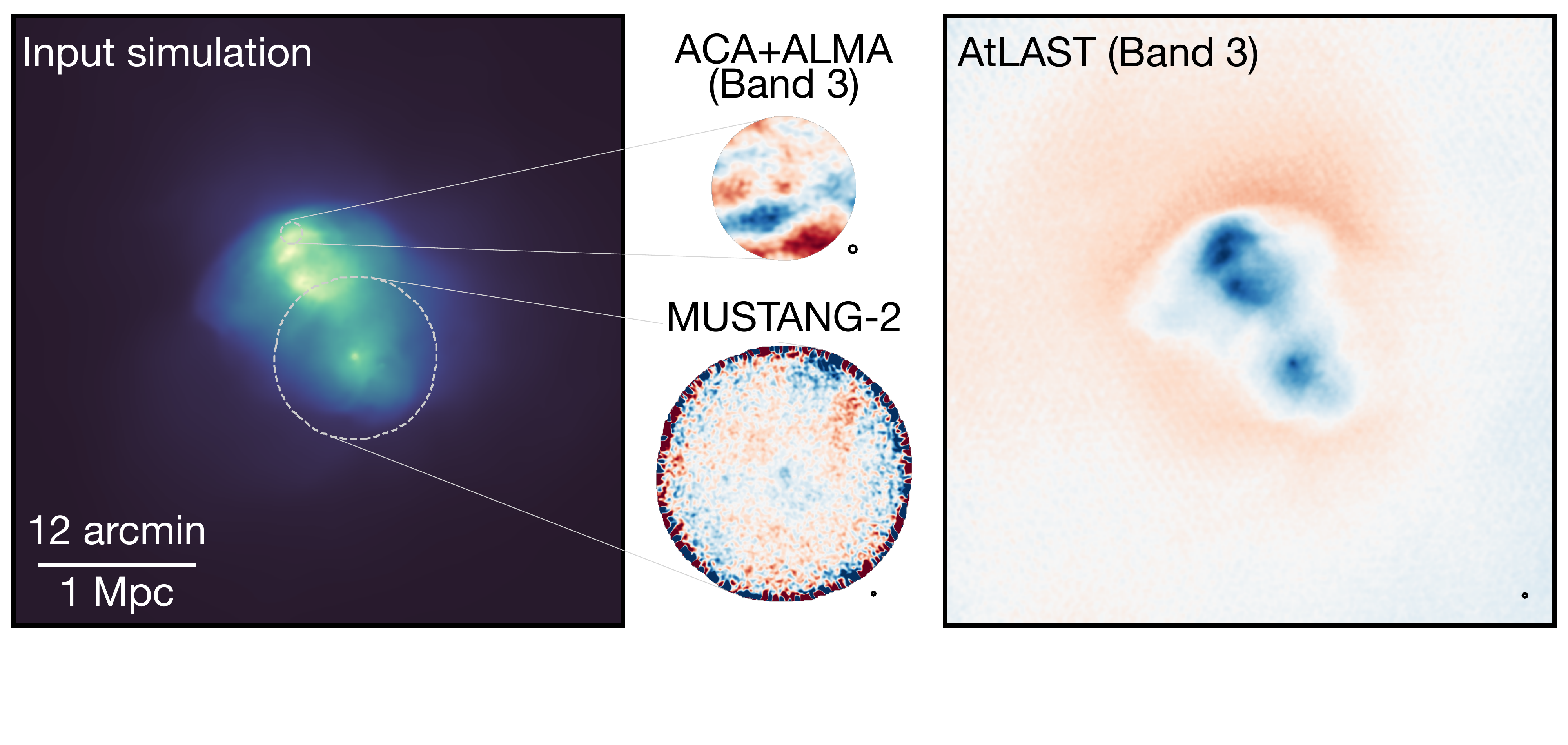}
    \vspace{-21pt}
    \caption{Simulated nearby galaxy cluster ($M_{500}\!\!=\!\!1.28\times10^{15}\,\mathrm{M_{\odot}}$, $z\!\!=\!\!0.07$; left) as observed by ALMA+ACA in Band 3 (top centre), MUSTANG-2 (bottom centre), and AtLAST at $90~\mathrm{GHz}$ (right). Adapted from Di Mascolo et al. \citep{DiMascolo2025}.\vspace{-20pt}}
    \label{fig:mocks}
\end{wrapfigure}

Despite these advances, current SZ facilities have yet produced an incomplete view of the warm/hot Universe. Present-day large-aperture telescopes and interferometers (such as ALMA) alike lack the spectral coverage, mapping speed, and field of view (Fig.~\ref{fig:mocks}) required to disentangle the thermal, kinetic, and relativistic SZ components across the full extent of galaxy clusters. On the other hand, wide-field CMB experiments face complementary challenges: their coarse angular resolution and the resulting confusion from Galactic foreground emission and extragalactic backgrounds limit their ability to resolve small-scale SZ structures, while current and future radio-interferometers, whilst offering enhanced angular resolution and spectral coverage, filter out emission on arcminute scales and, therefore, are insensitive to the diffuse SZ signal from the cluster outskirts, nearby groups, and large-scale intergalactic medium. As a result, even when combining various existing telescopes, we lack a holistic view of the most massive and thermally evolved systems.
\smallskip

\textbf{\sffamily Drivers for next-generation SZ observations.} In order to make a decisive leap forward in our understanding of the formation and evolution of large-scale structures, next-generation SZ observations must overcome the previously mentioned limitations and access regimes that have remained beyond our observational reach. This includes pushing the redshift frontier to characterize the thermodynamic state of protoclusters and massive forming haloes at $z\gtrsim2-4$, where the earliest phases of hot gas accretion and thermalization take place. Equally crucial is the ability to probe lower-mass and diffuse structures in the nearby Universe -- filaments, galaxy clusters and groups, circumgalactic haloes -- that collectively host a substantial fraction of the ionized baryons and provide a record of the gravitational and feedback processes that have shaped their growth but remain poorly constrained observationally. Furthermore, direct measurements of small-scale SZ fluctuations reveal the imprint of turbulence. Characterizing its statistical properties can yield robust constraints on the non-thermal pressure support in cluster atmospheres \citep{Gaspari2014} -- a leading driver of biases to cluster-based cosmological applications.
Also, identifying and probing pressure discontinuities will shed light on how shocks mediate the energy 
\hspace{-6pt}injection into forming haloes \citep{Ha2018}. Such a thermal perspective can be complemented with kinetic SZ observations \citep{Biffi2022}, directly probing gas motions associated with mergers, accretion flows, and large-scale dynamics. This census of the warm/hot baryons will reveal how gas is assembled, thermalized, and redistributed within and between structures, ultimately informing our understanding of the thermal evolution of the Universe.

\vspace{-4pt}
\section{Technical requirements}
Resolving the small-scale signatures of the key astrophysical processes driving the thermal evolution of the warm/hot baryons demands a facility with capabilities extending well beyond current and planned (sub-)mm observatories (see Fig.~\ref{fig:mocks}). A core requirement is a large collecting area: a 50\,m-diameter single-dish telescope provides the necessary sensitivity to faint, diffuse SZ emission while enabling diffraction-limited beams of a few arcseconds at the highest frequencies. Optimal SZ observations will need to rely on the combination of the large collecting area with wide-field capabilities ($\gtrsim2\deg^2$), ensuring high throughput and sensitivity across arcsecond-to-degree scales in a single pointing. This will require the integration of next-generation, multichroic large-format cameras comprising up to $\sim10^6$ detector elements, aiming at yielding mapping speed $>10^3$ and up to $10^5$ times faster than existing (sub-)mm\linebreak facilities. Further, a full decomposition of the thermal, kinetic, and relativistic SZ signals demands broad, simultaneous spectral coverage across the SZ decrement, null, and increment ($\sim30-950~\mathrm{GHz}$; Fig.~\ref{fig:spectra}). 
Multi-band observations, including at least 8 discrete continuum bands \cite{DiMascolo2025}, are essential for controlling contamination from the cosmic infrared and radio backgrounds, and from Galactic foregrounds.

All of these requirements converge in the proposed concept \cite{Mroczkowski2025} of the Atacama Large Aperture Submm Telescope (AtLAST). Its unprecedented combination of high mapping speed, angular resolution, sensitivity, and spectral coverage will enable the first truly comprehensive census of the warm/hot baryons across all relevant mass, redshift, and spatial scales. Such a facility would position ESO at the forefront of (sub-)mm cosmology and large-scale structure studies. Moreover, AtLAST is designed to accommodate multiple instruments, enabling science well beyond the SZ domain -- e.g., through comprehensive surveys of the cold gas and dust in galaxies residing in overdense environments \citep{Lee2024ORE,vanKampen2024ORE}. In this way, AtLAST would extend ESO’s scientific reach across the full thermal history of the warm/hot Universe and establish a flagship observatory fully aligned with the ambitions of the Expanding Horizons.

\medskip

\textbf{\sffamily References.}\printbibliography[heading=none]

@ARTICLE{Abdulla2019,
       author = {{Abdulla}, Zubair and {Carlstrom}, John E. and {Mantz}, Adam B. and {Marrone}, Daniel P. and {Greer}, Christopher H. and {Lamb}, James W. and {Leitch}, Erik M. and {Muchovej}, Stephen and {O'Donnell}, Christine and {Plagge}, Thomas J. and {Woody}, David},
        title = "{Constraints on the Thermal Contents of the X-Ray Cavities of Cluster MS 0735.6+7421 with Sunyaev-Zel{\textquoteright}dovich Effect Observations}",
      journal = {\apj},
         year = 2019,
        month = feb,
       volume = {871},
       number = {2},
          eid = {195},
        pages = {195},
          doi = {10.3847/1538-4357/aaf888},
archivePrefix = {arXiv},
       eprint = {1806.05050},
 primaryClass = {astro-ph.GA},
       adsurl = {https://ui.adsabs.harvard.edu/abs/2019ApJ...871..195A}
}

@ARTICLE{ACT2025,
       author = {{ACT-DES-HSC}},
        title = "{The Atacama Cosmology Telescope: DR6 Sunyaev-Zel'dovich Selected Galaxy Clusters Catalog}",
      journal = {arXiv},
         year = 2025,
        month = jul,
          eid = {2507.21459},
        pages = {2507.21459},
          doi = {10.48550/arXiv.2507.21459},
archivePrefix = {arXiv},
       eprint = {2507.21459},
 primaryClass = {astro-ph.CO},
       adsurl = {https://ui.adsabs.harvard.edu/abs/2025arXiv250721459A}
}

@ARTICLE{Adam2025,
       author = {{Adam}, R. and {Eynard-Machet}, T. and {Bartalucci}, I. and {Cherouvrier}, D. and {Clerc}, N. and {Di Mascolo}, L. and {Dupourqu{\'e}}, S. and {Ferrari}, C. and {Mac{\'\i}as-P{\'e}rez}, J. -F. and {Pointecouteau}, E. and {Pratt}, G.~W.},
        title = "{PITSZI: Probing intra-cluster medium turbulence with Sunyaev{\textendash}Zel'dovich imaging: Application to the triple merging cluster MACS J0717.5+3745}",
      journal = {\aap},
         year = 2025,
        month = feb,
       volume = {694},
          eid = {A182},
        pages = {A182},
          doi = {10.1051/0004-6361/202452342},
archivePrefix = {arXiv},
       eprint = {2409.14804},
 primaryClass = {astro-ph.CO},
       adsurl = {https://ui.adsabs.harvard.edu/abs/2025A&A...694A.182A}
}

@ARTICLE{Basu2016,
       author = {{Basu}, K. and {Sommer}, M. and {Erler}, J. and {Eckert}, D. and {Vazza}, F. and {Magnelli}, B. and {Bertoldi}, F. and {Tozzi}, P.},
        title = "{ALMA-SZ Detection of a Galaxy Cluster Merger Shock at Half the Age of the Universe}",
      journal = {\apjl},
         year = 2016,
        month = oct,
       volume = {829},
       number = {2},
          eid = {L23},
        pages = {L23},
          doi = {10.3847/2041-8205/829/2/L23},
archivePrefix = {arXiv},
       eprint = {1608.05413},
 primaryClass = {astro-ph.CO},
       adsurl = {https://ui.adsabs.harvard.edu/abs/2016ApJ...829L..23B}
}

@ARTICLE{Biffi2022,
       author = {{Biffi}, Veronica and {ZuHone}, John A. and {Mroczkowski}, Tony and {Bulbul}, Esra and {Forman}, William},
        title = "{The velocity structure of the intracluster medium during a major merger: Simulated microcalorimeter observations}",
      journal = {\aap},
         year = 2022,
        month = jul,
       volume = {663},
          eid = {A76},
        pages = {A76},
          doi = {10.1051/0004-6361/202142764},
archivePrefix = {arXiv},
       eprint = {2201.12370},
 primaryClass = {astro-ph.CO},
       adsurl = {https://ui.adsabs.harvard.edu/abs/2022A&A...663A..76B}
}

@ARTICLE{Bleem2023,
	journal={OJAp},
	doi={10.21105/astro.2311.07512},
	publisher={Maynooth Academic Publishing},
	title={Galaxy Clusters Discovered via the Thermal Sunyaev-Zel’dovich Effect in the 500-square-degree SPTpol Survey},
	volume=7,
	author={Bleem, L.E. and Klein, M. and Abbott, T. M. C. and Ade, P. A. R. and Aguena, M. and Alves, O. and Anderson, A. J. and Andrade-Oliveira, F. and Ansarinejad, B. and Archipley, M. and Ashby, M. L. N. and Austermann, J. E. and Bacon, D. and Beall, J. A. and Bender, A. N. and Benson, B. A. and Bianchini, F. and Bocquet, S. and Brooks, D. and Burke, D. L. and Calzadilla, M. and Carlstrom, J. E. and Rosell, A. Carnero and Carretero, J. and Chang, C. L. and Chaubal, P. and Chiang, H. C. and Chou, T-L. and Citron, R. and Moran, C. Corbett and Costanzi, M. and Crawford, T. M. and Crites, A. T. and da Costa, L. N. and de Haan, T. and De Vicente, J. and Desai, S. and Dobbs, M. A. and Doel, P. and Everett, W. and Ferrero, I. and Flaugher, B. and Floyd, B. and Friedel, D. and Frieman, J. and Gallicchio, J. and Garc'ia-Bellido, J. and Gatti, M. and George, E. M. and Giannini, G. and Grandis, S. and Gruen, D. and Gruendl, R. A. and Gupta, N. and Gutierrez, G. and Halverson, N. W. and Hinton, S. R. and Holder, G. P. and Hollowood, D. L. and Holzapfel, W. L. and Honscheid, K. and Hrubes, J. D. and Huang, N. and Hubmayr, J. and Irwin, K. D. and Mena-Fernández, J. and James, D. J. and Kéruzoré, F. and Knox, L. and Kuehn, K. and Lahav, O. and Lee, A. T. and Lee, S. and Li, D. and Lowitz, A. and Marshal, J. L. and McDonald, M. and McMahon, J. J. and Menanteau, F. and Meyer, S. S. and Miquel, R. and Mohr, J. J. and Montgomery, J. and Myles, J. and Natoli, T. and Nibarger, J. P. and Noble, G. I. and Novosad, V. and Ogando, R. L. C. and Padin, S. and Patil, S. and Pereira, M. E. S. and Pieres, A. and Malag'on, A. A. Plazas and Pryke, C. and Reichardt, C. L. and Rodr'iguez-Monroy, M. and Romer, A. K. and Ruhl, J. E. and Saliwanchik, B. R. and Salvati, L. and Sanchez, E. and Saro, A. and Schaffer, K. K. and Schrabback, T. and Sevilla-Noarbe, I. and Sievers, C. and Smecher, G. and Smith, M. and Somboonpanyakul, T. and Stalder, B. and Stark, A. A. and Suchyta, E. and Swanson, M. E. C. and Tarle, G. and To, C. and Tucker, C. and Veach, T. and Vieira, J. D. and Vincenzi, M. and Wang, G. and Weller, J. and Whitehorn, N. and Wiseman, P. and Wu, W. L. K. and Yefremenko, V. and Zebrowski, J. A. and Zhang, Y.},
	date={2024-02-09},
	year=2024,
	month=2,
	day=9,
          eid = {2311.07512},
        pages = {2311.07512},
archivePrefix = {arXiv},
       eprint = {2311.07512},
 primaryClass = {astro-ph.CO},
       adsurl = {https://doi.org/10.21105/astro.2311.07512}
}

@article{Bulbul2024,
       author = {{Bulbul}, E. and {Liu}, A. and {Kluge}, M. and {Zhang}, X. and {Sanders}, J.~S. and {Bahar}, Y.~E. and {Ghirardini}, V. and {Artis}, E. and {Seppi}, R. and {Garrel}, C. and {Ramos-Ceja}, M.~E. and {Comparat}, J. and {Balzer}, F. and {B{\"o}ckmann}, K. and {Br{\"u}ggen}, M. and {Clerc}, N. and {Dennerl}, K. and {Dolag}, K. and {Freyberg}, M. and {Grandis}, S. and {Gruen}, D. and {Kleinebreil}, F. and {Krippendorf}, S. and {Lamer}, G. and {Merloni}, A. and {Migkas}, K. and {Nandra}, K. and {Pacaud}, F. and {Predehl}, P. and {Reiprich}, T.~H. and {Schrabback}, T. and {Veronica}, A. and {Weller}, J. and {Zelmer}, S.},
        title = "{The SRG/eROSITA All-Sky Survey. The first catalog of galaxy clusters and groups in the Western Galactic Hemisphere}",
      journal = {\aap},
         year = 2024,
        month = may,
       volume = {685},
          eid = {A106},
        pages = {A106},
          doi = {10.1051/0004-6361/202348264},
archivePrefix = {arXiv},
       eprint = {2402.08452},
 primaryClass = {astro-ph.CO},
       adsurl = {https://ui.adsabs.harvard.edu/abs/2024A&A...685A.106B}
}

@ARTICLE{Chluba2012,
       author = {{Chluba}, Jens and {Nagai}, Daisuke and {Sazonov}, Sergey and {Nelson}, Kaylea},
        title = "{A fast and accurate method for computing the Sunyaev-Zel'dovich signal of hot galaxy clusters}",
      journal = {\mnras},
         year = 2012,
        month = oct,
       volume = {426},
       number = {1},
        pages = {510},
          doi = {10.1111/j.1365-2966.2012.21741.x},
archivePrefix = {arXiv},
       eprint = {1205.5778},
 primaryClass = {astro-ph.CO},
       adsurl = {https://ui.adsabs.harvard.edu/abs/2012MNRAS.426..510C}
}

@ARTICLE{deGraaff2019,
       author = {{de Graaff}, Anna and {Cai}, Yan-Chuan and {Heymans}, Catherine and {Peacock}, John A.},
        title = "{Probing the missing baryons with the Sunyaev-Zel'dovich effect from filaments}",
      journal = {\aap},
         year = 2019,
        month = apr,
       volume = {624},
          eid = {A48},
        pages = {A48},
          doi = {10.1051/0004-6361/201935159},
archivePrefix = {arXiv},
       eprint = {1709.10378},
 primaryClass = {astro-ph.CO},
       adsurl = {https://ui.adsabs.harvard.edu/abs/2019A&A...624A..48D}
}

@ARTICLE{DiMascolo2019,
       author = {{Di Mascolo}, Luca and {Mroczkowski}, Tony and {Churazov}, Eugene and {Markevitch}, Maxim and {Basu}, Kaustuv and {Clarke}, Tracy E. and {Devlin}, Mark and {Mason}, Brian S. and {Randall}, Scott W. and {Reese}, Erik D. and {Sunyaev}, Rashid and {Wik}, Daniel R.},
        title = "{An ALMA+ACA measurement of the shock in the Bullet Cluster}",
      journal = {\aap},
         year = 2019,
        month = aug,
       volume = {628},
          eid = {A100},
        pages = {A100},
          doi = {10.1051/0004-6361/201936184},
archivePrefix = {arXiv},
       eprint = {1907.07680},
 primaryClass = {astro-ph.CO},
       adsurl = {https://ui.adsabs.harvard.edu/abs/2019A&A...628A.100D}
}

@ARTICLE{DiMascolo2023,
       author = {{Di Mascolo}, Luca and {Saro}, Alexandro and {Mroczkowski}, Tony and {Borgani}, Stefano and {Churazov}, Eugene and {Rasia}, Elena and {Tozzi}, Paolo and {Dannerbauer}, Helmut and {Basu}, Kaustuv and {Carilli}, Christopher L. and {Ginolfi}, Michele and {Miley}, George and {Nonino}, Mario and {Pannella}, Maurilio and {Pentericci}, Laura and {Rizzo}, Francesca},
        title = "{Forming intracluster gas in a galaxy protocluster at a redshift of 2.16}",
      journal = {\nat},
         year = 2023,
        month = mar,
       volume = {615},
       number = {7954},
        pages = {809},
          doi = {10.1038/s41586-023-05761-x},
archivePrefix = {arXiv},
       eprint = {2303.16226},
 primaryClass = {astro-ph.CO},
       adsurl = {https://ui.adsabs.harvard.edu/abs/2023Natur.615..809D}
}

@ARTICLE{DiMascolo2025,
       author = {{Di Mascolo}, Luca and {Perrott}, Yvette and {Mroczkowski}, Tony and {Raghunathan}, Srinivasan and {Andreon}, Stefano and {Ettori}, Stefano and {Simionescu}, Aurora and {van Marrewijk}, Joshiwa and {Cicone}, Claudia and {Lee}, Minju and {Nelson}, Dylan and {Sommovigo}, Laura and {Booth}, Mark and {Klaassen}, Pamela and {Andreani}, Paola and {Cordiner}, Martin A. and {Johnstone}, Doug and {van Kampen}, Eelco and {Liu}, Daizhong and {Maccarone}, Thomas J. and {Morris}, Thomas W. and {Orlowski-Scherer}, John and {Saintonge}, Am{\'e}lie and {Smith}, Matthew and {Thelen}, Alexander E. and {Wedemeyer}, Sven},
        title = "{Atacama Large Aperture Submillimeter Telescope (AtLAST) science: Resolving the hot and ionized Universe through the Sunyaev-Zeldovich effect}",
      journal = {ORE},
         year = 2025,
        month = jun,
       volume = {4},
          eid = {113},
        pages = {113},
          doi = {10.12688/openreseurope.17449.2},
archivePrefix = {arXiv},
       eprint = {2403.00909},
 primaryClass = {astro-ph.CO},
       adsurl = {https://ui.adsabs.harvard.edu/abs/2025ORE.....4..113D}
}

@ARTICLE{Gaspari2014,
       author = {{Gaspari}, M. and {Churazov}, E. and {Nagai}, D. and {Lau}, E.~T. and {Zhuravleva}, I.},
        title = "{The relation between gas density and velocity power spectra in galaxy clusters: High-resolution hydrodynamic simulations and the role of conduction}",
      journal = {\aap},
         year = 2014,
        month = sep,
       volume = {569},
          eid = {A67},
        pages = {A67},
          doi = {10.1051/0004-6361/201424043},
archivePrefix = {arXiv},
       eprint = {1404.5302},
 primaryClass = {astro-ph.CO},
       adsurl = {https://ui.adsabs.harvard.edu/abs/2014A&A...569A..67G}
}

@ARTICLE{Gobat2019,
       author = {{Gobat}, R. and {Daddi}, E. and {Coogan}, R.~T. and {Le Brun}, A.~M.~C. and {Bournaud}, F. and {Melin}, J. -B. and {Riechers}, D.~A. and {Sargent}, M. and {Valentino}, F. and {Hwang}, H.~S. and {Finoguenov}, A. and {Strazzullo}, V.},
        title = "{Sunyaev-Zel'dovich detection of the galaxy cluster Cl J1449+0856 at z = 1.99: The pressure profile in uv space}",
      journal = {\aap},
         year = 2019,
        month = sep,
       volume = {629},
          eid = {A104},
        pages = {A104},
          doi = {10.1051/0004-6361/201935862},
archivePrefix = {arXiv},
       eprint = {1907.10985},
 primaryClass = {astro-ph.CO},
       adsurl = {https://ui.adsabs.harvard.edu/abs/2019A&A...629A.104G}
}

@ARTICLE{Ha2018,
       author = {{Ha}, Ji-Hoon and {Ryu}, Dongsu and {Kang}, Hyesung},
        title = "{Properties of Merger Shocks in Merging Galaxy Clusters}",
      journal = {\apj},
         year = 2018,
        month = apr,
       volume = {857},
       number = {1},
          eid = {26},
        pages = {26},
          doi = {10.3847/1538-4357/aab4a2},
archivePrefix = {arXiv},
       eprint = {1706.05509},
 primaryClass = {astro-ph.CO},
       adsurl = {https://ui.adsabs.harvard.edu/abs/2018ApJ...857...26H}
}

@ARTICLE{Hincks2022,
       author = {{Hincks}, Adam D. and {Radiconi}, Federico and {Romero}, Charles and {Madhavacheril}, Mathew S. and {Mroczkowski}, Tony and {Austermann}, Jason E. and {Barbavara}, Eleonora and {Battaglia}, Nicholas and {Battistelli}, Elia and {Bond}, J. Richard and {Calabrese}, Erminia and {de Bernardis}, Paolo and {Devlin}, Mark J. and {Dicker}, Simon R. and {Duff}, Shannon M. and {Duivenvoorden}, Adriaan J. and {Dunkley}, Jo and {D{\"u}nner}, Rolando and {Gallardo}, Patricio A. and {Govoni}, Federica and {Hill}, J. Colin and {Hilton}, Matt and {Hubmayr}, Johannes and {Hughes}, John P. and {Lamagna}, Luca and {Lokken}, Martine and {Masi}, Silvia and {Mason}, Brian S. and {McMahon}, Jeff and {Moodley}, Kavilan and {Murgia}, Matteo and {Naess}, Sigurd and {Page}, Lyman and {Piacentini}, Francesco and {Salatino}, Maria and {Sarazin}, Craig L. and {Schillaci}, Alessandro and {Sievers}, Jonathan L. and {Sif{\'o}n}, Crist{\'o}bal and {Staggs}, Suzanne and {Ullom}, Joel N. and {Vacca}, Valentina and {Van Engelen}, Alexander and {Vissers}, Michael R. and {Wollack}, Edward J. and {Xu}, Zhilei},
        title = "{A high-resolution view of the filament of gas between Abell 399 and Abell 401 from the Atacama Cosmology Telescope and MUSTANG-2}",
      journal = {\mnras},
         year = 2022,
        month = mar,
       volume = {510},
       number = {3},
        pages = {3335},
          doi = {10.1093/mnras/stab3391},
archivePrefix = {arXiv},
       eprint = {2107.04611},
 primaryClass = {astro-ph.CO},
       adsurl = {https://ui.adsabs.harvard.edu/abs/2022MNRAS.510.3335H}
}

@ARTICLE{Khabibullin2018,
       author = {{Khabibullin}, I. and {Komarov}, S. and {Churazov}, E. and {Schekochihin}, A.},
        title = "{Polarization of Sunyaev-Zel'dovich signal due to electron pressure anisotropy in galaxy clusters}",
      journal = {\mnras},
         year = 2018,
        month = feb,
       volume = {474},
       number = {2},
        pages = {2389},
          doi = {10.1093/mnras/stx2924},
archivePrefix = {arXiv},
       eprint = {1711.03084},
 primaryClass = {astro-ph.HE},
       adsurl = {https://ui.adsabs.harvard.edu/abs/2018MNRAS.474.2389K}
}

@ARTICLE{Khatri2016,
       author = {{Khatri}, Rishi and {Gaspari}, Massimo},
        title = "{Thermal SZ fluctuations in the ICM: probing turbulence and thermodynamics in Coma cluster with Planck}",
      journal = {\mnras},
         year = 2016,
        month = nov,
       volume = {463},
       number = {1},
        pages = {655},
          doi = {10.1093/mnras/stw2027},
archivePrefix = {arXiv},
       eprint = {1604.03106},
 primaryClass = {astro-ph.CO},
       adsurl = {https://ui.adsabs.harvard.edu/abs/2016MNRAS.463..655K}
}

@ARTICLE{Lee2024,
       author = {{Lee}, Elizabeth and {Chluba}, Jens},
        title = "{The SZ effect with anisotropic distributions and high energy electrons}",
      journal = {\jcap},
         year = 2024,
        month = jul,
       volume = {2024},
       number = {7},
          eid = {040},
        pages = {040},
          doi = {10.1088/1475-7516/2024/07/040},
archivePrefix = {arXiv},
       eprint = {2403.18530},
 primaryClass = {astro-ph.HE},
       adsurl = {https://ui.adsabs.harvard.edu/abs/2024JCAP...07..040L}
}

@ARTICLE{Lee2024ORE,
       author = {{Lee}, Minju and {Schimek}, Alice and {Cicone}, Claudia and {Andreani}, Paola and {Popping}, Gergo and {Sommovigo}, Laura and {Appleton}, Philip N. and {Bischetti}, Manuela and {Cantalupo}, Sebastiano and {Chen}, Chian-Chou and {Dannerbauer}, Helmut and {De Breuck}, Carlos and {Di Mascolo}, Luca and {Emonts}, Bjorn H.~C. and {Hatziminaoglou}, Evanthia and {Pensabene}, Antonio and {Rizzo}, Francesca and {Rybak}, Matus and {Shen}, Sijing and {Lundgren}, Andreas and {Booth}, Mark and {Klaassen}, Pamela and {Mroczkowski}, Tony and {Cordiner}, Martin A. and {Johnstone}, Doug and {van Kampen}, Eelco and {Liu}, Daizhong and {Maccarone}, Thomas and {Saintonge}, Amelie and {Smith}, Matthew and {Thelen}, Alexander E. and {Wedemeyer}, Sven},
        title = "{Atacama Large Aperture Submillimeter Telescope (AtLAST) science: The hidden circumgalactic medium}",
      journal = {ORE},
         year = 2024,
        month = jun,
       volume = {4},
        pages = {117},
          doi = {10.12688/openreseurope.17452.1},
archivePrefix = {arXiv},
       eprint = {2403.00924},
 primaryClass = {astro-ph.GA},
       adsurl = {https://ui.adsabs.harvard.edu/abs/2024ORE.....4..117L}
}

@ARTICLE{Liu2025,
       author = {{Liu}, R. Henry and {Ferraro}, Simone and {Schaan}, Emmanuel and {Zhou}, Rongpu and {Aguilar}, Jessica Nicole and {Ahlen}, Steven and {Battaglia}, Nicholas and {Bianchi}, Davide and {Brooks}, David and {Claybaugh}, Todd and {Cole}, Shaun and {Coulton}, William R. and {de la Macorra}, Axel and {Dey}, Arjun and {Fanning}, Kevin and {Forero-Romero}, Jaime E. and {Gazta{\~n}aga}, Enrique and {Gong}, Yulin and {Gontcho}, Satya Gontcho A. and {Gruen}, Daniel and {Gutierrez}, Gaston and {Hadzhiyska}, Boryana and {Honscheid}, Klaus and {Howlett}, Cullan and {Kehoe}, Robert and {Kisner}, Theodore and {Kremin}, Anthony and {Kusiak}, Aleksandra and {Lambert}, Andrew and {Landriau}, Martin and {Le Guillou}, Laurent and {Levi}, Michael and {Lokken}, Martine and {Manera}, Marc and {Martini}, Paul and {Meisner}, Aaron and {Miquel}, Ramon and {Moodley}, Kavilan and {Newman}, Jeffrey A. and {Niz}, Gustavo and {Palanque-Delabrouille}, Nathalie and {Percival}, Will and {Prada}, Francisco and {P{\'e}rez-R{\`a}fols}, Ignasi and {Ried Guachalla}, Bernardita and {Rossi}, Graziano and {Sanchez}, Eusebio and {Schlegel}, David and {Schubnell}, Michael and {Seo}, Hee-Jong and {Sif{\'o}n}, Crist{\'o}bal and {Sprayberry}, David and {Tarl{\'e}}, Gregory and {Vavagiakis}, Eve M. and {Weaver}, Benjamin Alan and {Wollack}, Edward J. and {Zou}, Hu},
        title = "{Measurements of the thermal Sunyaev-Zel'dovich effect with ACT and DESI luminous red galaxies}",
      journal = {\prd},
         year = 2025,
        month = oct,
       volume = {112},
       number = {8},
          eid = {083561},
        pages = {083561},
          doi = {10.1103/jqn8-19gx},
archivePrefix = {arXiv},
       eprint = {2502.08850},
 primaryClass = {astro-ph.CO},
       adsurl = {https://ui.adsabs.harvard.edu/abs/2025PhRvD.112h3561L}
}

\end{document}